# KEY MANGAMENT IN WIRELESS SENSOR NETWORKS USING A MODIFIED BLOM SCHEME

Rohith Singi Reddy
Computer Science Department
Oklahoma State University, Stillwater, OK 74078

**ABSTRACT**
Key establishment between any pair of nodes is an essential requirement for providing secure services in wireless sensor networks. Blom's scheme is a prominent key management scheme but its shortcomings include large computation overhead and memory cost. We propose a new scheme in this paper that modifies Blom's scheme in a manner that reduces memory and computation costs. This paper also provides the value for secure parameter t such that the network is resilient.

**INTRODUCTION**
Sensor networks comprise a number of sensors with limited resources that are used to collect environmental information and they have been considered for various purposes including security monitoring, target tracking and research activities in hazardous environments. Recent advances in both electronic and computer technologies have increased the demand of wireless sensor networks. Since authentication and confidentiality protocols require an agreed key between the nodes and security of the communication depends on the cryptographic schemes employed, key management is a very important security issue in wireless sensor networks. Earlier key management protocols were based on either symmetric or asymmetric cryptographic functions. Due to resource limitations and lack of security in the sensors, key management protocols based on public keys are inefficient. Hence, symmetric algorithm based key management protocols are used in wireless sensor networks [6].

Several people have proposed different approaches to address the problem of key management [8], [9], [10], [11], [12], [13]. The simple and easy method that could be employed is an online key management center [7]. But, this approach carries a high overhead. Another interesting method is to pre-distribute keys among the sensors, which can result in low cost key establishment in wireless sensor networks. But such schemes fail in handling security or efficiency problems. Some key management protocols use deployment knowledge to increase efficiency. In particular, key agreement schemes for wireless sensor networks must satisfy low energy consumption, low cost, low memory usage, lack of trusted infrastructure, resilient against node capture.

Eschenauer and Gligor [1] proposed probabilistic key pre-distribution to establish pair-wise keys between neighboring nodes. In their scheme, each node is preloaded with a key subset from a global key pool in such a way that any two neighboring nodes can share at least one common key with a certain probability. This scheme is vulnerable to the node compromise attack, where keys of normal nodes can be known when some nodes are compromised by adversaries. Chan, Perrig, and Song [2] proposed the q-composite random key pre-distribution scheme, in which they modified E-G scheme by only increasing the number of keys that two random nodes share from at least 1 to at least q. Their scheme achieves good security under small scale attacks while increased vulnerability in large scale node compromise attack. Later, Du et al [3] proposed the multiple-space key pre-distribution scheme where each key is replaced by a special key space and many more people came with certain modifications to the existing scheme. All these schemes assume a random node deployment model where each sensor node has direct pair-wise keys shared with only portion of the neighbors, and depends on the multi hop or the path which is established in order for the nodes to communicate with long distance nodes.

A different method of key management in sensor networks is given in [17]. This method requires marking the sensors in a ordered manner. Data in distributed networks can also be protected by using the method of partitioning [14]-[16], but that will not be considered here.

In this paper, we propose a new key management scheme that is based on the pre-distribution scheme of Blom [3]. This paper provides a modification to the original Blom's scheme and presents a solution to reduce computation overhead and memory cost. The modification is based on the use of the Hadamard matrix instead of the Vandermonde matrix of the Blom scheme.

**The Blom Scheme**
Blom's key distribution method [4] allows any pair of users in the system to find a unique shared key. According to this method, a network with N users and a collusion of less than t+1 users cannot reveal the keys which are held by other users. Thus the security of the network depends on the chosen value of t, which is called Blom's secure parameter (t<<N). Larger value of t leads to greater resilience but one needs to be careful in choosing a high value because that increases the amount of memory required to store key information.

During the initialization phase, a central authority or base station first constructs a (t + 1)×N matrix P over a finite field GF(q), where N is the size of the network and q is the prime number. P is known to all users and it can be constructed using Vandermonde matrix. It can be shown that any t+1 columns of P are linearly independent when $n_i$, i=1, 2,…N are all distinct.

$$P = \begin{bmatrix} 1 & 1 & 1 & \cdots & 1 \\ n_1 & n_2 & n_3 & \cdots & n_N \\ n_1^2 & n_2^2 & n_3^2 & \cdots & n_N^2 \\ \cdots\cdots\cdots\cdots\cdots\cdots\cdots\cdots\cdots \\ n_1^t & n_2^t & n_3^t & \cdots & n_N^t \end{bmatrix}$$

Then the central authority selects a random (t + 1) × (t + 1) symmetric matrix S over GF(q), where S is secret and only known by the central authority. An N ×(t + 1) matrix A = (S. P)$^T$ is computed. Because S is symmetric, it is easy to see

$$K = A.P = (S.P)^T. P = P^T. S^T.P$$
$$= P^T.S.P = (A.P)^T = K^T$$

User pair (i, j) will use $K_{ij}$, the element in row i and column j in K, as the shared key. Because $K_{ij}$ is calculated by the i-th row of A and the j-th column of P, the central authority assigns the i-th row of A matrix and the i-th column of P matrix to each user i, for 1, 2….. N. Therefore, when user i and user j need to establish a shared key between them, they first exchange their columns of P, and then they can compute $K_{ij}$ and $K_{ji}$, respectively, using their private rows of A. The t-secure parameter guarantees that no compromise of up to t nodes has any information about $K_{ij}$ or $K_{ji}$.

## PROPOSED SCHEME

**Modified Blom Scheme**
The proposed method makes use of the original Blom's scheme [4]. In the original Blom's scheme all the computations involved in generating the keys are based on Vandermonde matrix which is a public matrix (P) and known to even the adversaries. Here, to make sure that any t+1 columns of P are linearly independent i.e., to generate unique keys, all the values in the matrix are chosen to be distinct. However, for large values of t, number of rows in the matrix increases and which in turn corresponds to a greater value in the columns because the column values increases in a geometric series. In the Blom's scheme [4] for any two nodes to generate a common key, each node should store column of public matrix and row of the calculated secret matrix. Since every sensor node is provided with limited memory and energy, it will be difficult to store both the row and column in the sensor memory for a large network.

To reduce the computation and memory overhead in Blom's scheme, instead of using Vandermonde matrix [5] we propose the use of non-binary Hadamard matrix as the public matrix. As, the Hadamard matrix is a square matrix with 1s and -1s, it reduces of complexity of calculating values for all the elements corresponding to the columns in Vandermonde matrix. Another advantage of using Hadamard matrix is it reduces the cost of saving the columns in the memory of sensor because any node can easily generate Hadamard matrix of known size. Similar to Blom's scheme the operation which are to be performed to generate the keys will depend on the prime number i.e., the number which depends on the desired key length.

$$\begin{bmatrix} 1 & 1 & 1 & 1 \\ 1 & -1 & 1 & -1 \\ 1 & 1 & -1 & -1 \\ 1 & -1 & -1 & 1 \end{bmatrix}$$

Figure 1. Binary Hadamard Matrix

$$\begin{bmatrix} 1 & 1 & 1 & 1 \\ 1 & 30 & 1 & 30 \\ 1 & 1 & 30 & 30 \\ 1 & 30 & 30 & 1 \end{bmatrix}$$

Figure 2. Non-Binary Hadamard Matrix for Modulo 31

The original binary form Haramard matrix is depicted in Figure 1. The only change that is made to the Hadamard matrix is all negative numbers are replaced with the non-negative prime number. We can observe that the Hadamard matrix shown in Figure 2 contains equal numbers of one's and the prime number, so all the further calculations will be very simple. Key generation technique is similar to that used in the Blom's scheme [4]. Following are the steps involved in calculating the key.

(1) Initially N × N form of the Hadamard matrix is considered and depending on the t value first t rows along with N columns are selected as the public matrix. The construction of Hadamard matrix depends on the prime number (q), we must set q to be larger than the network size (q>N).

(2) The central authority selects a random $(t + 1) \times (t + 1)$ symmetric matrix S, where S is secret and only known by the central authority. An $N \times (t + 1)$ matrix $A = (S \cdot P)^T$ is computed.

(3) The central authority stores each row of the matrix A in the node memory with corresponding index. This is shown in Figure 3.

(4) Finally user pair (i, j) can compute the key by generating the Hadamard matrix and multiplying the secret row stored in the node with column of the Hadamard matrix corresponding to the node index with which it want to communicate. The key generation between any two nodes is shown in Figure 4.

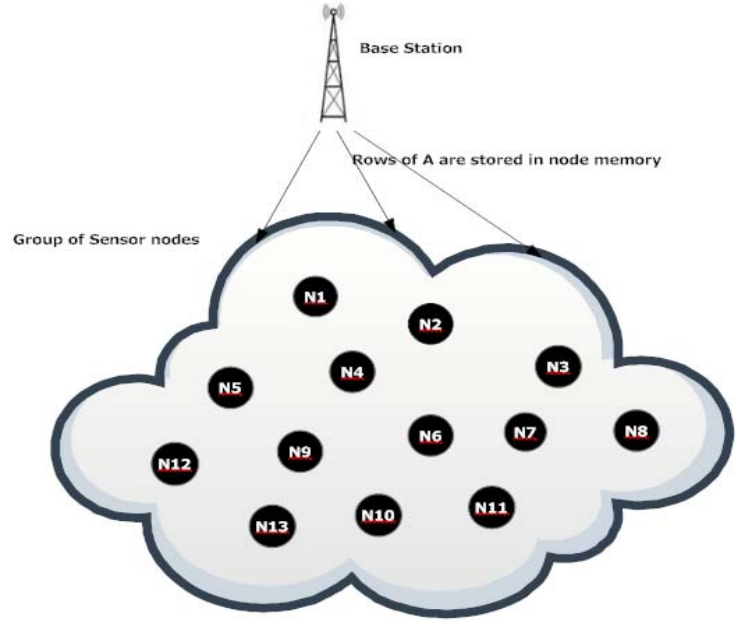

Figure 3. Storing rows into node memory

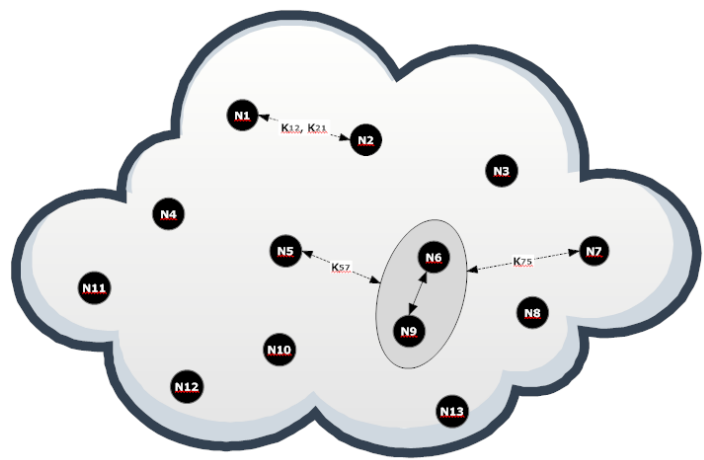

Figure 4. Key Establishment between any pair of nodes

By integrating the proposed scheme along with the schemes which previously relied on Blom's scheme such as Du. et al. [3] we can gain significant memory utilization along with reduced computations.

**Example**

The following example shows the working of the modified Blom's scheme using Hadamard matrix. Let the number of nodes in the network be 8, secure parameter t=6 and prime number q=31 which says if no more than 6 nodes in the network are compromised it is not possible to find the keys of other users. The assumption is that all the sensor nodes are within the transmission range other and can communicate directly.

**Modified Hadamard Matrix**

$$\begin{bmatrix} 1 & 1 & 1 & 1 & 1 & 1 & 1 & 1 \\ 1 & 30 & 1 & 30 & 1 & 30 & 1 & 30 \\ 1 & 1 & 30 & 30 & 1 & 1 & 30 & 30 \\ 1 & 30 & 30 & 1 & 1 & 30 & 30 & 1 \\ 1 & 1 & 1 & 1 & 30 & 30 & 30 & 30 \\ 1 & 30 & 1 & 30 & 30 & 1 & 30 & 1 \\ 1 & 1 & 30 & 30 & 30 & 30 & 1 & 1 \\ 1 & 30 & 30 & 1 & 30 & 1 & 1 & 30 \end{bmatrix}$$

**Public Matrix (P)**

$$P = \begin{bmatrix} 1 & 1 & 1 & 1 & 1 & 1 & 1 & 1 \\ 1 & 30 & 1 & 30 & 1 & 30 & 1 & 30 \\ 1 & 1 & 30 & 30 & 1 & 1 & 30 & 30 \\ 1 & 30 & 30 & 1 & 1 & 30 & 30 & 1 \\ 1 & 1 & 1 & 1 & 30 & 30 & 30 & 30 \\ 1 & 30 & 1 & 30 & 30 & 1 & 30 & 1 \end{bmatrix}$$

Let secret matrix (S) is any symmetric matrix.

$$S = \begin{bmatrix} 3 & 11 & 15 & 28 & 7 & 5 \\ 11 & 30 & 4 & 1 & 2 & 8 \\ 15 & 4 & 6 & 14 & 18 & 21 \\ 28 & 1 & 14 & 17 & 25 & 6 \\ 7 & 2 & 18 & 25 & 27 & 9 \\ 5 & 8 & 21 & 6 & 9 & 8 \end{bmatrix}$$

A= (S.P)$^T$

$$S.P = \begin{bmatrix} 69 & 1345 & 1316 & 968 & 417 & 1403 & 1664 & 1026 \\ 56 & 1187 & 201 & 1274 & 346 & 1013 & 491 & 1100 \\ 78 & 1209 & 658 & 977 & 1209 & 1122 & 1789 & 890 \\ 91 & 787 & 990 & 700 & 990 & 1338 & 1889 & 1251 \\ 88 & 1132 & 1335 & 929 & 1132 & 1654 & 2379 & 1451 \\ 57 & 695 & 840 & 1130 & 550 & 724 & 1333 & 1159 \end{bmatrix} \text{mod } 31$$

$$A = (S.P)^T = \begin{bmatrix} 7 & 25 & 16 & 29 & 26 & 26 \\ 12 & 9 & 0 & 12 & 16 & 13 \\ 14 & 15 & 7 & 29 & 2 & 3 \\ 7 & 3 & 16 & 18 & 30 & 14 \\ 14 & 5 & 0 & 29 & 16 & 23 \\ 8 & 21 & 6 & 5 & 11 & 11 \\ 21 & 26 & 22 & 29 & 23 & 0 \\ 3 & 15 & 22 & 11 & 25 & 12 \end{bmatrix}$$

After calculating matrix A each sensor node memory is filled with unique row chosen from matrix A corresponding to the same index. It shows that all the computations involved in calculating the matrix A are simple.

**Generating Key**
Any pair of nodes must generate a key for securely communicating with each other. Since, the Hadamard matrix is easy to construct, the overhead of storing a column in node memory is reduced. Suppose node two and node eight want to communicate with each other, first they generate the column of neighboring node using the Hadamard matrix and then multiply it with the row stored in the node memory.

$$K_{2,8} = A_2.P_8 = \begin{bmatrix} 12 & 9 & 0 & 12 & 16 & 13 \end{bmatrix} \begin{bmatrix} 1 \\ 30 \\ 30 \\ 1 \\ 30 \\ 1 \end{bmatrix} = 787 \bmod 31 = 12$$

$$K_{8,2} = A_8.P_2 = \begin{bmatrix} 3 & 15 & 22 & 11 & 25 & 12 \end{bmatrix} \begin{bmatrix} 1 \\ 30 \\ 1 \\ 30 \\ 1 \\ 30 \end{bmatrix} = 1190 \bmod 31 = 12$$

We can observe that both nodes generate a common key and further communication between them will make use of the pair-wise key.

In general matrix K can be represented as shown below and we can notice that the symmetric nature gives the same key for any pair of nodes such that $K_{ij} = K_{ji}$.

K= A.P

$$K = \begin{bmatrix} 5 & 0 & 8 & 26 & 25 & 0 & 28 & 26 \\ 0 & 25 & 7 & 18 & 4 & 19 & 11 & 12 \\ 8 & 7 & 29 & 20 & 29 & 9 & 19 & 22 \\ 26 & 18 & 20 & 22 & 0 & 17 & 25 & 21 \\ 25 & 4 & 29 & 0 & 9 & 18 & 13 & 14 \\ 0 & 19 & 9 & 17 & 18 & 19 & 27 & 17 \\ 28 & 11 & 19 & 25 & 13 & 27 & 4 & 10 \\ 26 & 12 & 22 & 21 & 14 & 17 & 10 & 26 \end{bmatrix}$$

# SIMULATION

The Blom's scheme provides a rough estimation for the value of secure parameter t and it doesn't give any desired range for the parameter such that the entire network is secure within the range. The proposed scheme gives the value for secure parameter t so that all the nodes in the network can communicate with other in a secure manner.

**Secure Parameter (t)**

Simulations are carried on with a network size of 16, 32 and 64 with varying key length. Each network is tested with different value of the parameter t. It is observed that for any network we can generate large number of unique keys if the parameter size chosen is greater than half the size of the network. In general for a network with size N, the value of t should be t>=N/2+1

It is observed that the prime number chosen to compute the keys has a great impact on number of unique keys generated in the network. For minimum prime number it shows to have maximum number of unique keys at N/2+1 but total number of unique keys increases if the value of prime number is increased.

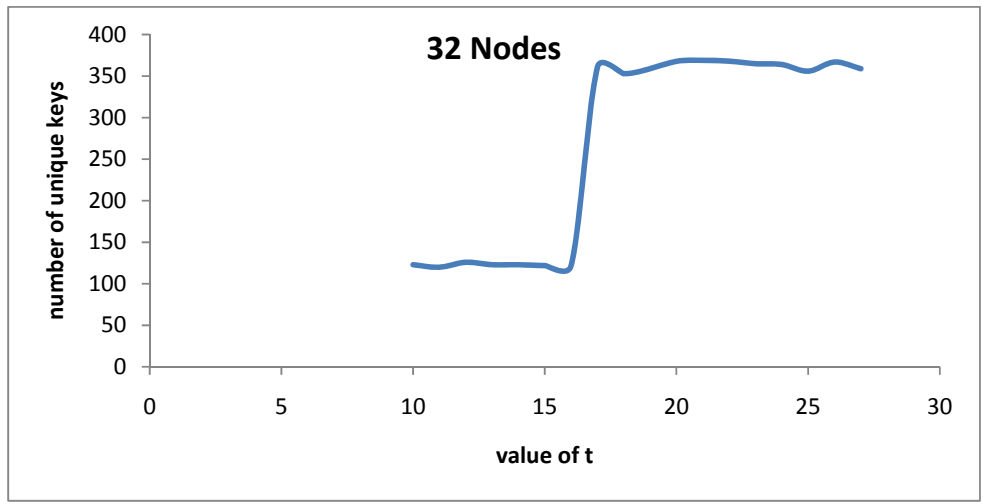

Figure 4 Number of unique keys for a network of 32 nodes

Figure 4 shows value of t versus number of unique keys for a network with 32 nodes and for prime number 751. It is observed that for a t value greater the half the size of network (t=17) total number of unique keys generated by the network has increased drastically. Moreover, there is not much difference is observed if the t value is increased further. Figure 5 shows value of t versus number of unique keys for a network with 64 nodes and for prime number 1181. Similar observation is made and found that total number of unique keys increases drastically at t=33. Perhaps, the total number of unique keys varies for different network sizes and different prime numbers.

It is observed that for any network containing N nodes if the value of t is chosen as t=N/2 +1 then as long as t+1 nodes are compromised the remaining uncompromised nodes will communicate in a secure manner i.e., the network is resilient at t=N/2+1.

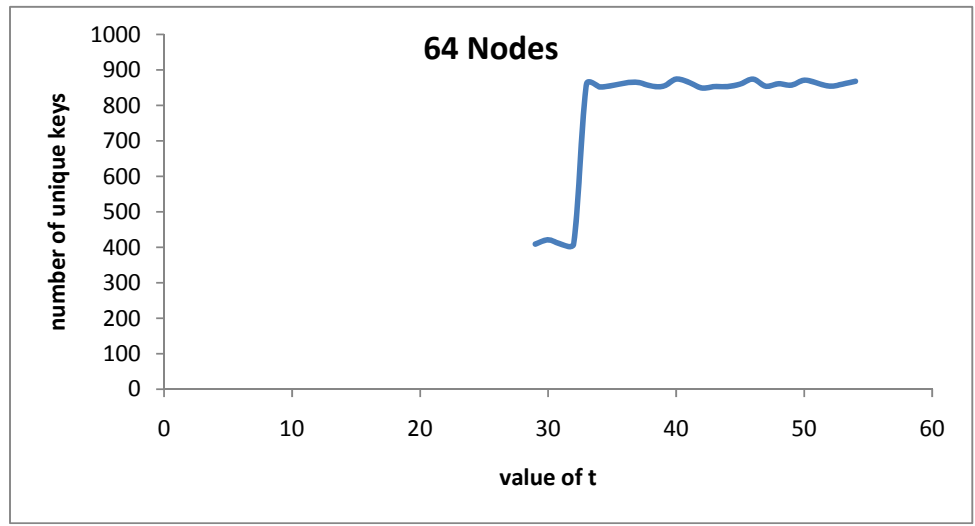

Figure 5 Number of unique keys for a network of 64 nodes

**CONCLUSION**

This paper presents an alternative way of generating the keys using the Blom's scheme. The proposed method has the advantage of reducing the computation overhead and memory costs over the original Blom's scheme. In addition, this scheme provides the minimum value for the secure parameter t such that the network is more resilient. It may be noted that other decomposition schemes such as LU, LQ, QR, may be used to generate a key between any pair of nodes.